\documentclass{elsart}
\usepackage{graphics}
\usepackage{subeqnarray}
\usepackage{epsfig}
\usepackage{txfonts}
\usepackage{cancel}
\usepackage{natbib}
\usepackage{amssymb}
\begin{document}
\begin{frontmatter}
\title{Drug absorption through a cell monolayer: a theoretical work on a non-linear three-compartment model}
\author{Niko Komin}\footnote{Corresponding author.  {\sl Email address}: {\tt niko@ifisc.uib-csic.es}, Tel: +34-971-259520, fax: +34-971-173426},
\author{Ra\'ul Toral}
\address{IFISC (Instituto de F{\'\i}sica Interdisciplinar y Sistemas Complejos), UIB-CSIC, Campus UIB, 07122 Palma de Mallorca,Spain}
\date{\today}
\begin{abstract}
The subject of analysis is a non-linear three-compartment model, widely used in pharmacological absorption studies. It has been transformed into a general form, thus leading automatically to an appropriate approximation. This made the absorption profile accessible and expressions for absorption times, apparent permeabilities and equilibrium values were given. These findings allowed a profound analysis of results from non-linear curve fits and delivered the dependencies on the systems' parameters over a wide range of values. The results were applied to an absorption experiment with multidrug transporter-affected antibiotic CNV97100 on Caco-2 cell monolayers.
\end{abstract}


\end{frontmatter}

\section{Introduction}

\begin{figure}
 \centering
   \includegraphics[width=10cm]{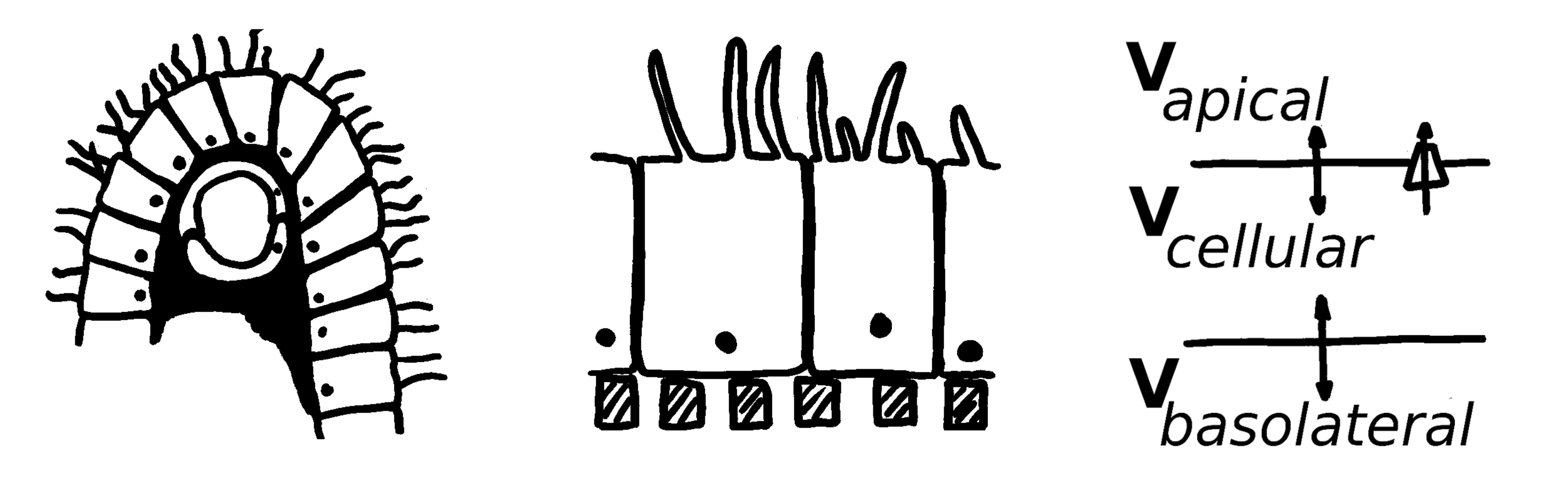}
 \caption{schematic view of: intestinal wall with {\it microvilli} and a capillary embedded in the {\it lamina propria} (left); epithelial cells on filter (middle); simplified mathematical model with efflux pump indicated on apical side (right)}
 \label{fig:intestine}
\end{figure}

Orally administered drugs are mainly absorbed by the small intestine; they are mediated upon by a variety of processes \citep{HunterHirst1997}.  The drug passes from the lumen through the epithelial cells and the {\it lamina propria} into the blood stream (fig. \ref{fig:intestine} (left)). On its way it can be metabolised, transported away from the tract where absorption is possible or accumulate in organs other than those of treatment. 

Much experimental activity aimed at analysing the kinetic aspects of the process of drug absorption has been pursued recently. For better control, a variety of {\it in-vitro} methods of drug absorption have been developed \citep{Balimane2000}. Epithelial cell cultures can be seeded in a mono-layer, forming the contact surface between two chambers (fig. \ref{fig:intestine} (middle)) and concentrations of an applied drug can be measured over time in both chambers. Two of the well-known cell culture models are Caco-2 cells \citep{ArturssonBorchardt1997,ArturssonPalm2001} and MDCK cells \citep{Irvine1999}.

Apart from a fundamental interest in understanding the basic mechanisms by which a drug is assimilated by the human body, the kinetics of drug absorption is also a topic of much practical interest. Detailed knowledge of this process, resulting in the prediction of the drug absorption profile, can be of much help in drug development stage~\citep{Eddershaw2000,Zhou2003}. To this end, several kinetic models for drug absorption within the body have been established (see e.g. \citep{YuAmidon1999}). In this paper we analyse in detail a previously developed model, belonging to the category of the so-called three-compartment models \citep{Skinner1959,Kramer1974} in which substances move between three volumes (e.g. the human organs) and active pumps are modelled by terms of non-linear fluxes. We will provide an analytical solution to an (adequate) approximation of the model, which facilitates analysis of the absorption characteristics as a whole without requiring repetitive numerical integration of the differential equations. This facilitates a fast and easy insight into how physiologically meaningful parameters influence quantities available from experiment in three-compartment models.

The paper is organised as follows: After introducing the model and presenting its solution in section \ref{sec:theModel}, we discuss the results and apply them to an antibiotic absorption study carried out with the compound CNV97100 on Caco-2 cell cultures (section \ref{sec:applicationResults}). Conclusions are found in section \ref{sec:conclusions}. Details of the mathematical procedure are summarised in the Appendix.

\section{The model and its theoretical treatment}
\label{sec:theModel}

Compartment models describe the behaviour of solutions or emulsions in connected volumes by analysing the molecule flux between them and all sources and sinks. When applied to drug absorption some specific simplifications must be made: Here it is considered that two volumes (e.g. gastrointestinal lumen and blood plasma {\it in-vivo} or apical and basolateral chamber {\it in-vitro}) are connected through a third, {\it in-vitro}: cellular, volume. The absorbed substance should have low lipophilicity such that the unstirred water layer can be neglected. Furthermore it is assumed that the compound does not ionise and that the concentrations in the different intestinal cells of the mono-layer are equal (which is exact only if all cells have the same parameters). With these assumptions the absorption can be seen as a transport from one large volume to another through a third (the cellular) volume \citep{Bermejo2005}. Figure \ref{fig:intestine} (right) sketches the simplifications of the model. There is no spatial dependency and molecules can pass through the two cell membranes. The overall amount of drug molecules is considered to be constant, a hypothesis that assumes a closed system and that metabolism does not occur. For an {\it in-vitro} experiment of short duration this is a reasonable assumption.

Passive transport across the membrane is mediated, to a first approximation, by the concentration gradients according to {\sc Fick}'s law \citep{Fick1855}, which specifies a linear relation between the flux of particles and the concentration gradient. When passive absorption is accompanied by energy-consuming efflux transporters, it is represented by a non-linear function term in kinetic transport equations. A variety of transporter types could be involved in the absorption of the molecules. In our work we consider that the non-linear transporters are present only on one (the apical) cell membrane, but our results could be extended directly to the case that those transporters are located on the basolateral membrane (or even in both membranes). Incorporating both linear and non-linear terms, the time evolution of the amount of diluted molecules ($Q_{A/C/B}$) in the three compartments can be described as follows:
\begin{subeqnarray}
  \label{eq:ThreeCompMod}
  \slabel{eq:ThreeCompMod1}
  \frac{dQ_A(t)}{dt}& = &\displaystyle
  +Cl_{AC}\left(\frac{Q_C}{V_C}-\frac{Q_A}{V_A}\right)+ {\cal J}\\
  \slabel{eq:ThreeCompMod2}
  \frac{dQ_C(t)}{dt}& = &\displaystyle -Cl_{AC}\left(\frac{Q_C}{V_C}-\frac{Q_A}{V_A}\right)-  {\cal J}-Cl_{CB}\left(\frac{Q_C}{V_C}-\frac{Q_B}{V_B}\right)\\
  \slabel{eq:ThreeCompMod3}
  \frac{dQ_B(t)}{dt}& = &\displaystyle  
  +Cl_{CB}\left(\frac{Q_C}{V_C}-\frac{Q_B}{V_B}\right)\\\nonumber\\
  \slabel{eq:ThreeCompMod4}
 Q_0&=&Q_A+Q_B+Q_C,
\end{subeqnarray}
where equation (\ref{eq:ThreeCompMod4}) stands for conservation of the overall molecule number, $Q_0$. The indices denote the corresponding compartment ({\it {\bf A}pical, {\bf C}ellular, {\bf B}asolateral}), $V_{A/C/B}$ are the respective volumes. The apical, cellular and basolateral concentrations are given respectively by $a=Q_A/V_A$, $c=Q_C/V_C$ and $b=Q_B/V_B$. The passive, linear, diffusion terms are proportional to the concentration difference, being $Cl_{AC}$ and $Cl_{CB}$ the clearances indexed with their respective membrane index. In the equations, ${\cal J}$ represents the non-linear contribution due to specific efflux transporters. As it is an energy-consuming process, this can happen both along or against the gradient.

In the appendix we will introduce an equivalent way of writing these kinetic equations (\ref{eq:ThreeCompMod}) that will demonstrate its formal similarity with some problems in the field of mechanics. This mechanical analogy will invite some approximations, consisting in a linearisation of (\ref{eq:ThreeCompMod}), allowing us to find explicit solutions for the evolution of the number of molecules on each compartment. If approximated this way, any mass-conserving, three-compartment-model will result in a sum of three exponentials, independent from the exact form of non-linearity:
\begin{subeqnarray}
\label{eq:solution_QaQbQc}
\slabel{eq:solution_Qa}
Q_A(t)&=&Q_A^{eq}-A_1 {\rm e}^{-t/t_1}-A_2 {\rm e}^{-t/t_2}-A_3 {\rm e}^{-t/t_3} \\
\slabel{eq:solution_Qb}
Q_B(t)&=&Q_B^{eq}-B_1 {\rm e}^{-t/t_1}-B_2 {\rm e}^{-t/t_2}-B_3 {\rm e}^{-t/t_3} \\
\slabel{eq:solution_Qc}
Q_C(t)&=&Q_C^{eq}-C_1 {\rm e}^{-t/t_1}-C_2 {\rm e}^{-t/t_2}-C_3 {\rm e}^{-t/t_3} 
\end{subeqnarray}
where $t_1$, $t_2$ and $t_3$ define three time-scales and $Q_A^{eq}$, $Q_B^{eq}$ and $Q_C^{eq}$ are the equilibrium asymptotic quantities of the diluted substance in each compartment. These and the constants $(A/B/C)_{(1/2/3)}$ adopt different expressions, depending on the non-linearity. If given a specific expression for the non-linear transport terms and numerical values for the parameters, one can calculate the above constants and compare the result with the numerical integration of the non-linear system (\ref{eq:ThreeCompMod}) or with experimental data. This explicit type of solution for a {\sc Michaelis-Menten} flux {\cal J} constitutes one of the main results of this paper and is the basis for the subsequent analysis. In section \ref{sec:applicationResults} we will carry out this program explicitly for the model and data taken from \citep{Bermejo2005}, resulting in parameters as given in table \ref{tab:coefficients2}.

Experimentally, it is rare to measure the complete variation of $Q_A(t)$, $Q_B(t)$ and $Q_C(t)$ with respect to time. A typical experiment \citep{Lentz2000,RuizGarcia2002,Faassen2003,Balimane2004,Bermejo2005}  starts by placing an initial concentration $C_0$ of a drug in either the apical or in the basolateral compartment. The so-called apparent permeability $\displaystyle P^{app}=\frac {dQ/dt}{S C_0}$, with $Q(t)$ the amount of material on the receiving side, is measured in both directions and the values are compared. 

The explicit solution Eqs.(\ref{eq:solution_QaQbQc}) identifies three different {\bf characteristic time scales}, $t_1$, $t_2$ and $t_3$, within the evolution of these concentrations. Each one of them separates well defined regimes in the evolution of the concentrations: if time is much smaller than the characteristic time scale, the corresponding exponential term comes close to being linear; it changes exponentially at times close to it and is almost constant at much larger times. This information helps the experimenter to decide if the chosen sampling interval is adequate or not. Furthermore, it is mathematically possible to observe oscillatory behaviour in the presented system, if one of the $t_{1,2,3}$ is complex. The condition for this to happen is shown in the appendix.

We would like to stress that our way of approximating the problem has allowed us to identify these natural time scales and to find their relationship to other constants that are experimentally accessible. It appears that in many cases (one example in the next section and table \ref{tab:ConcTauSlope}) one time-scale ($t_2$) is much smaller than the interval between measurements. Measurements of the apparent permeability are usually carried out within a time frame of between 15-30 min and a couple of hours \citep{Bermejo2005,Lentz2000,Yamashita2000}, whereas $t_2$ seems to be of the order of a few minutes. Hence, measurement times satisfy $t\gg t_2$ and the exponential term ${\rm e}^{-t/t_2}$ can be neglected. Analysis of the experiments \citep{Bermejo2005} indicates that transport is mediated by transporters with an intracellular binding site. In this case, and according to table \ref{tab:ConcTauSlope}, both $t_1$ and $t_3$ are much larger than the measurement times, allowing us to perform the linear approximations ${\rm e}^{-t/t_1}\approx 1-t/t_1$ and ${\rm e}^{-t/t_3}\approx 1-t/t_3$ to obtain explicit expressions for the apparent permeability:
\begin{equation}
\label{eq:Papp_BA}
P^{app}_{BA}=\frac{1}{S C_0}\left(\frac{A_1}{t_1}+\frac{A_3}{t_3}\right)
\end{equation}
in the case that the drug is initially delivered in the basolateral side, and
\begin{equation}
\label{eq:Papp_AB}
P^{app}_{AB}=\frac{1}{S C_0}\left(\frac{B_1}{t_1}+\frac{B_3}{t_3}\right)
\end{equation}
when the drug is delivered in the apical side. It seems that even in the case of very fast-absorbing drugs (tested for permeabilities such as those in \citep{Korjamo2007}) times are well separated into those of around an hour and those of less than a minute (data not shown). However, when it occurs that the time scales are all of a similar order, one can easily extend eqs. \ref{eq:Papp_BA} and \ref{eq:Papp_AB} by the required term.

Absorption of many drugs~\citep{RavivPollard1990} is seriously limited by P-glyco\-protein (P-gp), the multidrug transporter. This particular protein is expressed on the apical membrane of intestinal epithelium cells~\citep{TroutmanThakker2003A,TroutmanThakker2003B,RuizGarcia2002}. The molecule to be transported must bind with the protein and will then be ``flipped'' \citep{HunterHirst1997} onto the other side of the membrane, where it is no longer available for the ``reaction''. This makes its dynamics similar to enzyme reactions and is often represented by the sigmoid shape of a {\sc Michaelis-Menten}-reaction rate~\citep{MichMent1913}: 
\begin{eqnarray}
  \label{eq:MichMent}
  {\cal J}(Q_C)=\frac{S V_M Q_C/V_C}{K_M+Q_C/V_C} \;.
\end{eqnarray}
$V_M$ determines the maximal reaction velocity, $S$ is the surface area and $K_M$ is the concentration for which the velocity extends halfway towards the maximum. In this case of an {\bf intracellular binding} site, the relevant variable is the concentration $Q_C/V_C$ around the binding site of the transporter, inside the cell. 

In other cases, the efflux pump possesses an {\sl \bf extracellular binding} site. Consequently, transport is determined by the drug concentration, $Q_A/V_A$, in the apical compartment and the corresponding {\sc Michaelis-Menten} expression is:
\begin{eqnarray}
  \label{eq:MichMente}
  {\cal J}(Q_A)=\frac{S V_M Q_A/V_A}{K_M+Q_A/V_A} \;.
\end{eqnarray}

Both situations will be considered in this paper. For reasons of simplicity, we have considered that the efflux pumps depend on concentration on one of the two sides of the membrane \citep{RuizGarcia2002,Bermejo2005}. However, new results suggest that the transporter binding site for the molecule is inside the inner leaflet of the membrane \citep{hennessySpiers2006}. If we consider the space inside the phospholipid bilayer to be an additional volume with two permeable walls on either side, the concentration in that volume would be between those in the adjacent volumes. 

In the following section, these approximate expressions will be compared with the experimental data from \citep{Bermejo2005}.

\section{Results}
\label{sec:applicationResults}

\begin{figure}
 \centering
 \includegraphics{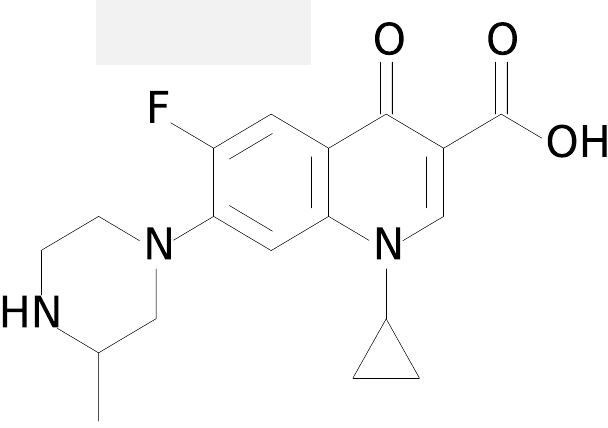}
 \caption{Structure of antibiotic CNV97100}
 \label{fig:cnv97100}
\end{figure}

Our mathematical treatment of the evolution equations has provided us with explicit expressions for the evolution over time of the amounts $Q_{A/B/C}(t)$, Eqs.(\ref{eq:solution_QaQbQc}),and the apparent permeability, Eqs.(\ref{eq:Papp_BA}) and (\ref{eq:Papp_AB}). In this section we intend to determine the validity of our approach. Firstly, by comparing theoretical predictions with the full results of numerical simulations of the evolution equations, we will show that our approximate treatment has a range of validity that covers typical experimental situations. 

Next, once this validity has been established, we will use our approximation to make comparisons to experimental data from an antibiotic (CNV97100, see fig.~\ref{fig:cnv97100}) absorption study \citep{Bermejo2005}.
In this experiment a $pH$ of 7.0 was used in both the apical and the basolateral chamber to avoid any bias due to ionisation effects. If the $pH$ were different in the two chambers it would be necessary to include these effects into the model by estimating the fraction of ionised and non ionised compound and to model separately the permeation of both species in both directions in each chamber. The unstirred water layer had not been considered to be the limiting diffusion step, an assumption justified by taking into account the molecular weight and the lipophilicity of CNV97100. This had been checked experimentally in situ in rats~\citep{Bermejo1999} and in vitro in Caco-2 cells~\citep{Bermejo2004}. As a consequence the three compartment model with a Michaelis-Menten-type flow was chosen to be an adequate picture of the underlying processes.

Finally, we will determine the variation of the {\it efflux ratio}, the equilibrium concentrations and the characteristic time scales for absorption with the system parameters.

\subsection {Concentration evolution and binding site location}

\begin{figure}
  \begin{center}
    \includegraphics[height=5cm]{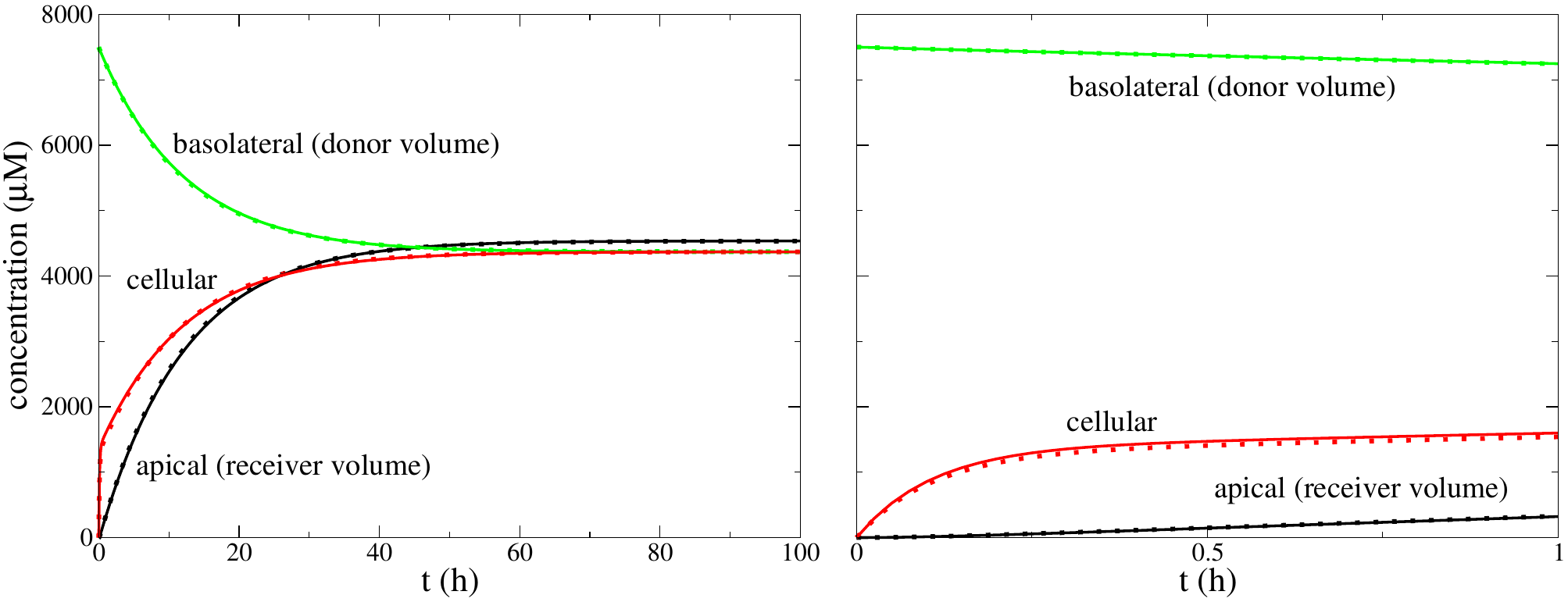}
    \caption{Time evolution of concentrations on either side of the cells and inside. {\it Dotted} line: Numerical integration of Eq. (\ref{eq:ThreeCompMod}). {\it Continuous} line: explicit solution Eq.(\ref{eq:solution_QaQbQc}). Parameters taken from \citep{Bermejo2005}, intracellular binding site with {\sc Michaelis-Menten} dynamics~(\ref{eq:MichMent}) is considered. Initial concentration $C_0=7500\mu M$ is applied on the basolateral side. {\it Right} graph: first hour amplified.} 
    \label{fig:timeline_basoLoad7500}
  \end{center}
\end{figure}

\begin{figure}
  \begin{center}
    \includegraphics[height=5cm]{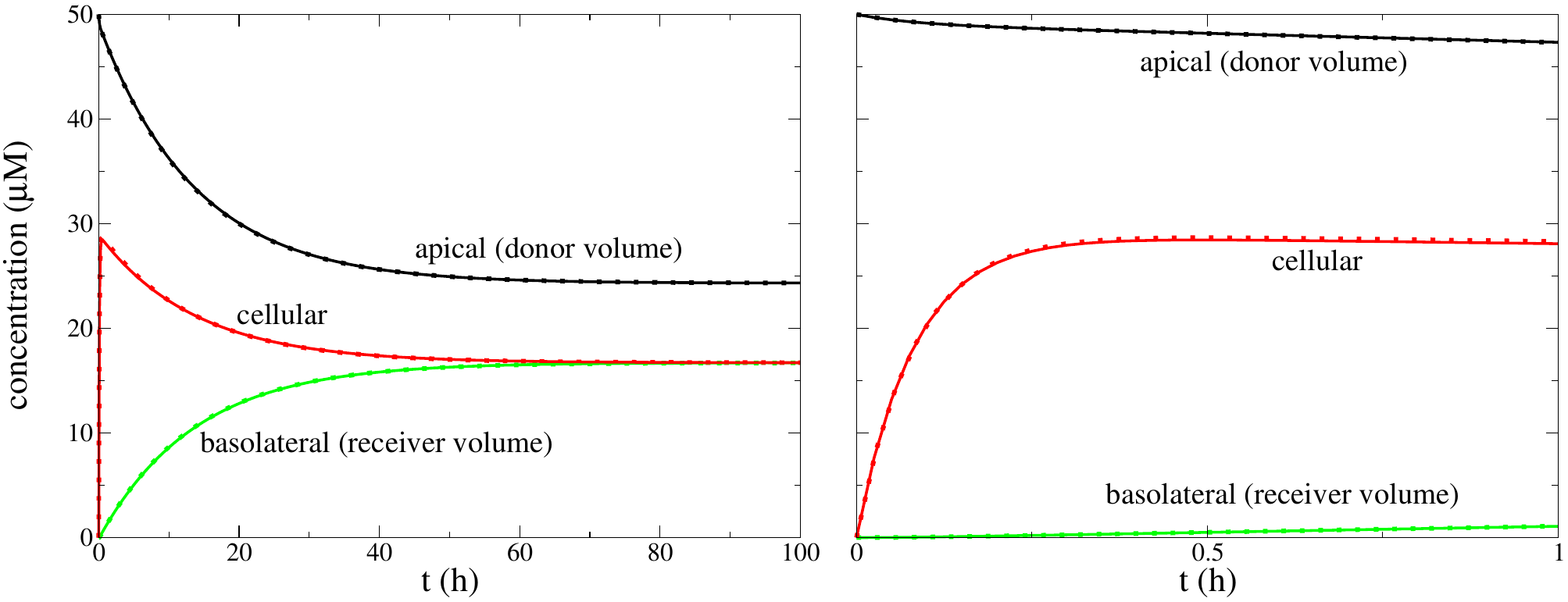}
    \caption{Same as figure \ref{fig:timeline_basoLoad7500}, but initial concentration $50\mu M$ applied apically.} 
    \label{fig:timeline_apicLoad0050}
  \end{center}
\end{figure}

We obtained expressions for the equilibrium amounts $Q_{A/B/C}^{eq}$ and the constants $(A/B/C)_{(1/2/3)}$ for non-linear transporters of the {\sc Michaelis-Menten} type, in cases of transport mediated by both intracellular and extracellular binding. The corresponding expressions are summarised in table \ref{tab:coefficients2} of the Appendix as a function of parameters of the model. Using the numerical values of those parameters as derived in the absorption study of reference \citep{Bermejo2005}, we can extract precise numerical values for the equilibrium amounts $Q_{A/B/C}^{eq}$ and the constants $(A/B/C)_{(1/2/3)}$. Those numerical values are also listed in table \ref{tab:coefficients2}. Finally, the numerical values of the time constants $t_{1/2/3}$ are listed in table \ref{tab:ConcTauSlope}. Using these numerical values, extracted from a real experiment and thus corresponding to a case of interest, we now proceed to check the accuracy of our approximation. To this end, we plotted in figures \ref{fig:timeline_basoLoad7500} and \ref{fig:timeline_apicLoad0050} (for two different initial conditions) the results of the direct numerical integration\footnote{For this numerical integration we used a fourth-order Runge-Kutta algorithm with a time step of $1s$.} of Eqs.(\ref{eq:ThreeCompMod}) and our approximation, Eqs.(\ref{eq:solution_QaQbQc}). The most noticeable feature is that, at the scale of the figures, the two approaches are nearly indistinguishable and, in fact, the difference is of the order of the thickness of the lines. We conclude that our treatment provides a simple, yet very precise, expression for the evolution of the amounts $Q_{A/B/C}(t)$ and can be used instead of the less transparent numerical integration of the equations. 

Table \ref{tab:ConcTauSlope} lists the apparent permeabilities and the resulting {\it efflux ratios}, $P^{app}_{BA}/P^{app}_{AB}$, as obtained from Eqs.(\ref{eq:Papp_BA}-\ref{eq:Papp_AB}) in the cases of internal and external binding site and different initial concentrations. The analysis of the three time scales $t_{1/2/3}$, listed in the same table, shows that  $t_2$ is well below the time of the first measurement (30min) for all cases studied and the duration of experiment ($t\sim 2$~h) satisfies $t\ll t_1,t_3$, hence validating the approximations that led to Eqs.(\ref{eq:Papp_BA}-\ref{eq:Papp_AB}). The same analysis justifies the validity of the linear fit used in the experimental studies to extract the apparent permeabilities from the data. To avoid overloading the paper with too many results, we have omitted the time scales for basolateral loading since they are of similar order. 

\begin{figure}
 \centering
 \includegraphics{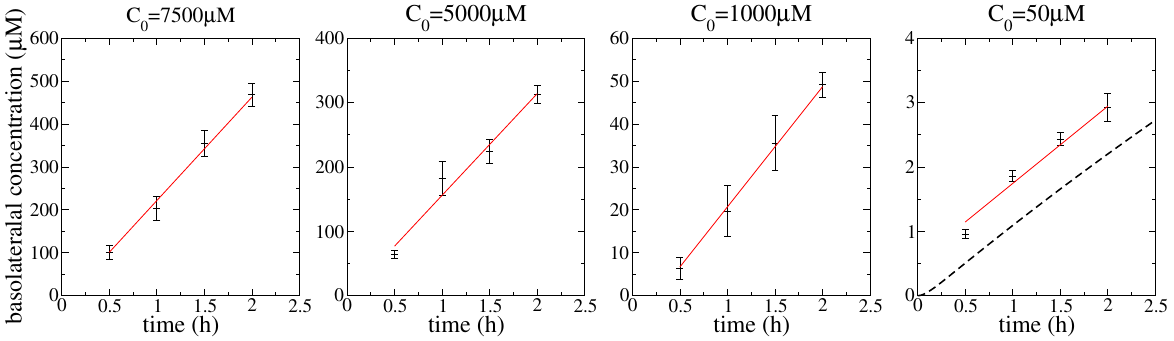}
 \includegraphics{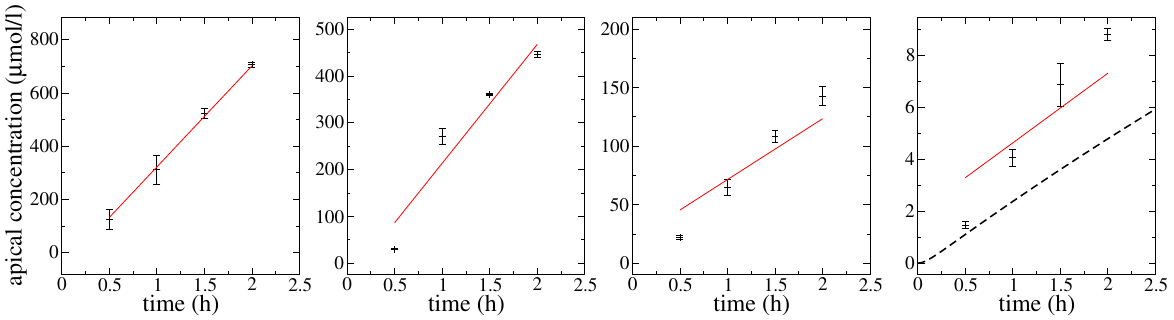}
 \caption{Antibiotic's concentration on the receiving side. ({\it Top}: drug loaded in apical compartment, {\it bottom}: basolateral loading.) Experimental values are from CNV97100 study. The {\it solid} line's slope is the prediction from the theoretical solution (shown in table \ref{tab:ConcTauSlope}) for intracellular binding, which was considered to be the model of best fit. The {\it dashed} line in the graphs on the right shows the numerical integration of the full system (\ref{eq:ThreeCompMod}).}
 \label{fig:experimentalData}
\end{figure}

To end the comparison with experimental data, we plot the results of the CNV97100 study in figure \ref{fig:experimentalData}, superimposing on the data a line with a slope equal to the apparent permeability from table \ref{tab:ConcTauSlope} (multiplied by $SC_0$).  If the antibiotic is loaded apically (top row in fig. \ref{fig:experimentalData}) the prediction of the model is good for all initial concentrations. If the loading is basolateral (bottom row); the prediction for low initial concentration underestimates the measured slope. As a consequence, the predicted efflux ratio differs from experiment at lower concentrations, which makes a more detailed analysis of this difference necessary. The {\sc Michaelis-Menten} kinetics used seem to provide an insufficient representation of P-gp efflux at lower initial concentrations when loaded basolaterally. The model underestimates the pump's efficiency in a basolateral to apical set up. We stress that our approximate solution still yields very accurate results for the apparent permeabilities and that this observed difference is a direct consequence of the model or the parameters used. To make this point clear, we have plotted the result of the numerical integration of (\ref{eq:ThreeCompMod}) for the lowest concentration of $50\mu M$ in figure \ref{fig:experimentalData}. The deviation from the experimental data is clearly observable in the case of basolateral loading of the drug. Considerations of other efflux pathways in the P-gp transporter protein are found for example in \citep{Acharya2006}.


\begin{table}[h]
        \caption{Time scales (when loaded apically) and apparent permeabilities predicted for different initial concentrations and different models (internal/external binding site).}
\label{tab:ConcTauSlope}
        \begin{center}
        \begin{tabular}{|cc|r@{.}lr@{.}lr@{.}l|r@{.}lr@{.}lr@{.}l|}
        \hline
$C_0$ & binding  & \multicolumn{2}{c}{$t_1$}& \multicolumn{2}{c}{$t_2$}& \multicolumn{2}{c|}{$t_3$}&
\multicolumn{2}{c}{$P^{app}_{BA}$}(cm s$^{-1}$)&
\multicolumn{2}{c}{$P^{app}_{AB}$}(cm s$^{-1}$)&
\multicolumn{2}{c|}{$\displaystyle \frac{P^{app}_{BA}}{P^{app}_{AB}}$}\\
\hline
7500&int &12&3h &6&38min &23&6h &6&70$\times 10^{-6}$  &6&37$\times 10^{-6}$ &1&05\\
    &ext &11&9h &6&67min &0&567h &6&80$\times 10^{-6}$  &4&12$\times 10^{-6}$ &1&65\\
\hline
5000&int &12&5h &6&27min &23&6h &6&73$\times 10^{-6}$  &6&26$\times 10^{-6}$ &1&07\\
    &ext &11&9h &6&67min &0&567h &6&86$\times 10^{-6}$  &3&97$\times 10^{-6}$ &1&73\\
\hline
1000&int &13&5h &5&59min &23&6h &6&88$\times 10^{-6}$  &5&57$\times 10^{-6}$ &1&24\\
    &ext &12&5h &6&70min &0&567h &7&18$\times 10^{-6}$  &3&10$\times 10^{-6}$ &2&32\\
\hline
50&int   &13&8h &4&90min &23&6h &7&10$\times 10^{-6}$  &4&88$\times 10^{-6}$ &1&45\\
    &ext &15&7h &6&73min &0&567h &6&94$\times 10^{-6}$  &2&07$\times 10^{-6}$ &3&35\\
\hline
  \end{tabular}
        \end{center}
\end{table}

\subsection{Parameter dependence}

\begin{figure}[t]
        \begin{center}
	\includegraphics[width=13cm]{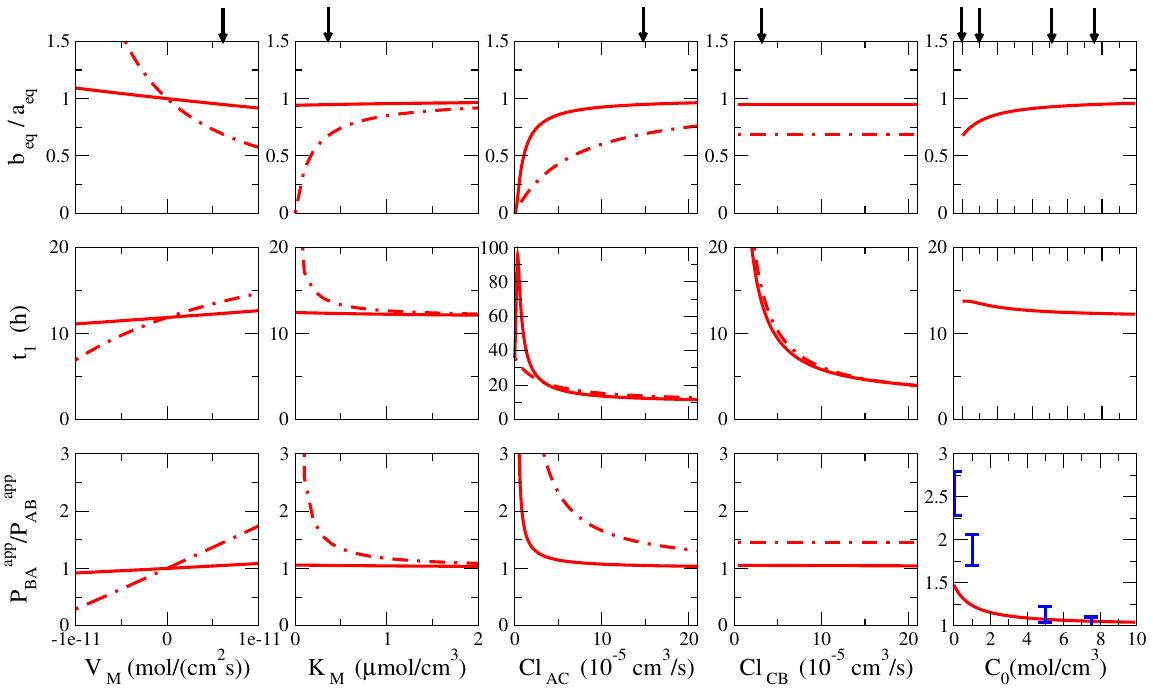}
        \caption{Secretory pump, intracellular binding site - {\it Top} and {\it middle}: equilibrium concentration ratio (basolateral/apical) and characteristic time (both for apical loading), {\it bottom}: efflux ratio $P^{app}_{BA}/P^{app}_{AB}$. Dependence on $V_M$ and $K_M$ (transporter parameters) and clearances $Cl_{AC/CB}$. {\it Continuous} line: $C_0=7500\mu M$, {\it dot-dashed} line: $C_0=50\mu M$. On the very {\it right}: dependence on initial concentration $C_0$. Arrows on top mark the experimentally derived value. (Values for the respectively fixed parameters taken from table \ref{tab:parameters}.) }
        \label{fig:parameter}
        \end{center}
\end{figure}

Once we had determined the validity of our approach, we wished to use the explicit expressions to determine the dependence on the system parameters of some quantities which are of experimental interest. Here lies one of the strengths of our solution: In figure \ref{fig:parameter} we plot the characteristic time scale for absorption $t_1$, the equilibrium concentration ratio on both cell sides $b_{eq}/a_{eq}$ and the efflux ratio $P^{app}_{BA}/P^{app}_{AB}$ as a function of the clearances $Cl_{AC}$ and $Cl_{CB}$, the pump parameters $V_M$ and $K_M$ and the initial concentration of drug $C_0$. Analytic formulas give access to these results much easier than repetitive integration throughout parameter space plus extracting the data from the resulting trajectories.

Again, for reasons of simplicity, we have limited our presentation to the case of a secretory pump located apically with intracellular binding site; the best model according to analysis by \citep{Bermejo2005}. As observed in this figure, an increase in the initial concentration $C_0$ implies a decrease in the characteristic time $t_1$ from a finite value to a minimum value, limiting $t_1$ to a certain range. Raising $C_0$ increases the equilibrium concentration ratio $b_{eq}/a_{eq}$.  Although this ratio varies significantly, steady concentration in the basolateral site, $b_{eq}$, shows a good linear dependence with $C_0$ (not shown in the figure). Note that the efflux ratio $P^{app}_{BA}/P^{app}_{AB}$ also decreases with increasing initial concentration, a feature supported by the experimental data, although the theoretical values deviate from the experimental results at low concentrations, a fact already discussed in the previous section. The clearance $Cl_{CB}$ of the membrane where the pump  is not situated has no influence on the equilibrium concentration and efflux ratios, but an increase of $Cl_{CB}$ decreases the characteristic time $t_1$, indicating a faster transport of the drug. On the other hand, an increase in the clearance $Cl_{AC}$ of the cell membrane where the pump is located has the effect of decreasing the efflux ratio and increasing the equilibrium concentration ratio. For large initial concentrations, $C_0=7500\mu M$, the characteristic time $t_1$ shows interesting behaviour with $Cl_{AC}$ since it first increases and then decreases, indicating very slow drug absorption for some intermediate values of the clearance.

At large concentrations, the three quantities analysed show small dependence with respect to the pump parameters $V_M$ and $K_M$, since the corresponding curves are almost flat. This makes it difficult to extract from the data accurate values of the pump parameters at those large concentrations. This suggests that lower concentrations would allow for a better experimental determination of the pump parameters - a practise used by experimentalists - however, we have to take into account, as discussed above, that the accuracy of the model might worsen with decreasing concentration. In the graph, we have included negative values for $V_M$, which is equivalent to a change in the flow direction of the pump.

Apart from these considerations, analysis of parameter dependency is the first step towards examining the propagation of errors into the experimentally available quantities. For example, it is clear from figure \ref{fig:parameter} (third and fourth column), that small differences in $Cl_{AC}$ would be nearly unnoticed, due to the rather flat curve around its measured value (marked by the black arrow on top of the figures). On the other hand, a small change in $Cl_{CB}$ yields a big variation of time scale $t_1$. More detailed investigations of errors and how they bias the results have been left aside for subsequent investigations.

\section{Discussion}
\label{sec:conclusions}
Three-compartment models are widely used in drug absorption studies. The one treated here consists of a linear part, representing passive absorption (for example through cell membranes), and a non-linear part which models other means of transport, such as ABC-transporter proteins embedded into the cell membrane. In this work we have shown a way to transform this model into the picture of a {\it ball in a well} (fig. \ref{fig:stoneInPotential}), which is a general form of equations facilitating analysis of mathematical structure. The lowest point of the ``well'' gives the value of the {\bf equilibrium concentration} after saturation of the process. With an adequate approximation, one can derive the {\bf absorption profile} as the sum of three exponentials and identify their {\bf characteristic time scales} dividing the process into phases of linear change, non-linear fluxes and saturation. In the phase of a near-linear profile we can provide explicit expressions for the {\bf apparent permeability}, a quantity usually measured in experiments.

These are general results which we used on a showcase system, where the non-linear transport is described by a {\sc Michaelis-Menten} profile - a common model of transporter proteins \citep{Sharma2002,VolkSchneider2003,Mizuarai2004}.  The approach presented in this paper make efflux ratio and time scales accessible. Both {\it apical} and {\it basolateral} drug loading can be treated by changing the initial conditions; we furthermore considered the possibilities of {\it intra}cellular as well as {\it extra}cellular binding sites. We analysed the importance of each physiological parameter in a wide range of values.

The presented results may contribute to a better understanding of the absorption process and to explaining differing observations in identical experimental setups from a more fundamental basis. Knowledge of parameter dependencies, a fundamental analysis of errors (confidence interval) and their consequences becomes possible.
This has been left aside for later studies.  The most promising experimental setups can furthermore be predicted by treating a newly proposed model in the same way.

\section*{Acknowledgements}

We would like to thank Marival Bermejo, Vicente Casabo, Isabel Gonz\'alez-\'Alvarez and other members of the group at the Pharmaceutical Department of Valencia University for continuous discussions, provision to us of the experimental data and their critical reading of an earlier version of this paper, as well as their hospitality on our visits to their laboratory. We acknowledge financial support from the EU NoE BioSim, LSHB-CT-2004-005137, and project FIS2007-60327 from MEC (Spain) and FEDER (EU). NK is supported by a grant from the Govern Balear.

\section{Appendix}
We will now give an overview of the used mathematical methods used and show the explicit solutions in the following subsections.

The conservation law Eq.~(\ref{eq:ThreeCompMod4}) reduces the number of independent variables from three to two concentrations. In other words, the system described by first-order differential equations (\ref{eq:ThreeCompMod}) only has two degrees of freedom. This implies that it can be replaced by a single differential equation of second order of the form:
\begin{equation}
	\label{eq:PotFricGeneral}
	\ddot{x}=-\Gamma\left(x\right)\dot{x} + F(x).
\end{equation}
Variable $x$ represents a rescaled concentration in one of the compartments (the cellular compartment if ${\cal J}$ depends on $Q_C$, as when using (\ref{eq:MichMent}), and the apical compartment if ${\cal J}$ depends on $Q_A$, as when using (\ref{eq:MichMente})); the speed, $\dot x$, and acceleration, $\ddot x$, are, respectively, the first and second derivatives of $x$ with respect to a rescaled time $s$, and $\Gamma(x)$ and $F(x)$ are functions to be described below. 

A qualitative understanding of these dynamics can be had by acknowledging that the previous equation corresponds to the equation of motion ({\sc Newton}'s second law) for the position $x$ of a particle of unit mass upon which a force $F(x)$ and a friction, proportional to the particle speed $\dot x$ act. The force $F(x)$ and the friction coefficient $\Gamma(x)$ contain all of the parameters of the system as well as the particular form of the non-linear flux (eqs. (\ref{eq:gamma-force})). 

\begin{figure}
 \centering
 \includegraphics[width=5.1cm]{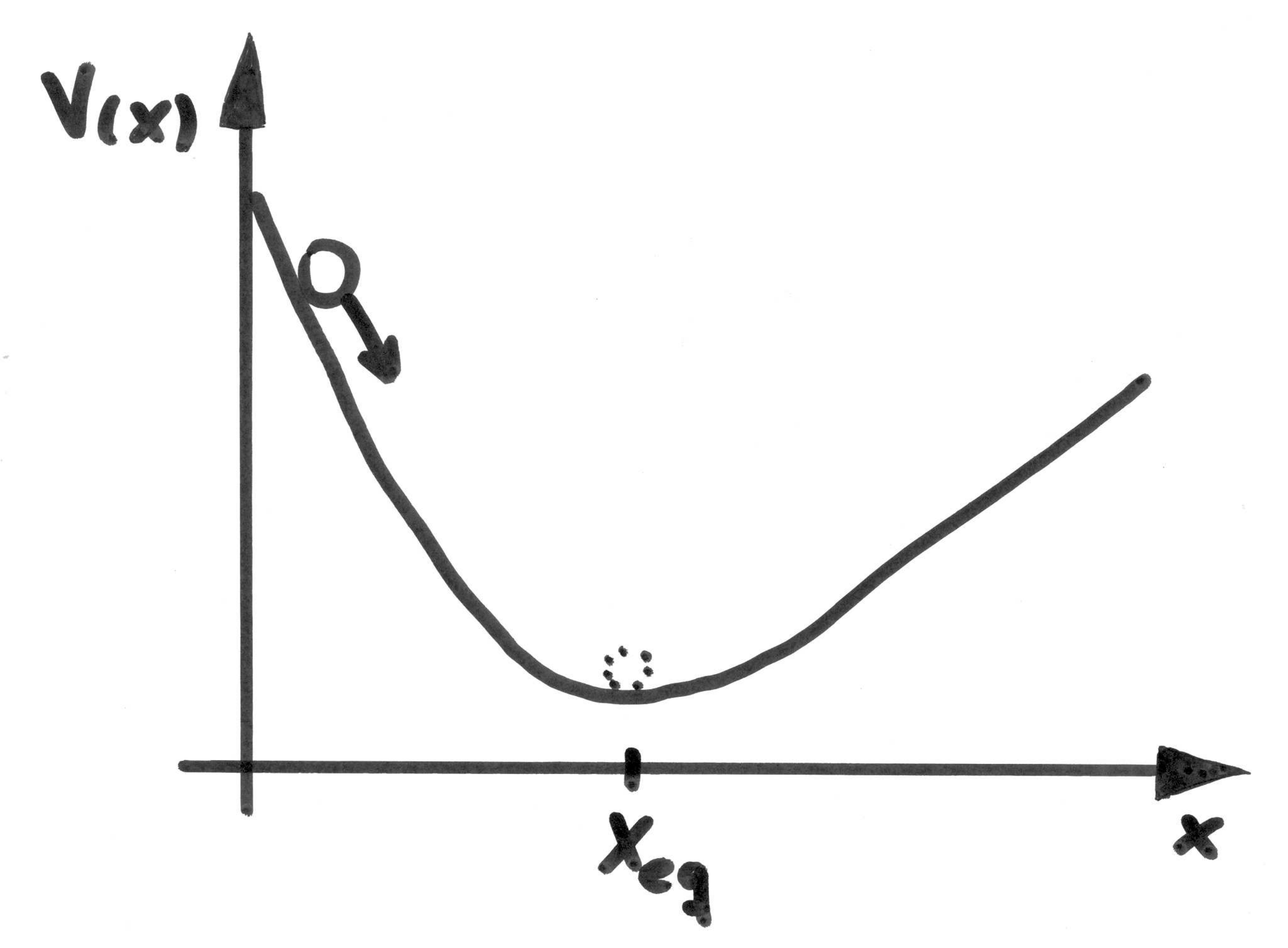}
	\caption{Visualisation of the particle-potential picture. The position of the ``ball'' represents the concentration which will eventually end up in the minimum.}
	\label{fig:stoneInPotential}
\end{figure}

If we now introduce the potential function $V(x)$ from which the force derives as $\displaystyle F(x)=-\frac{dV}{dx}$, the evolution of $x$ can be visualised as the relaxation of a ball rolling downwards within a well of shape $V(x)$ under the combined effects of gravity and friction. It is known from mechanics that the particle will eventually stop at the minimum of the potential (stable equilibrium state) and that relaxation towards this final state will proceed via damped oscillations or monotonously, depending on the relative strength of the friction and potential contributions. Figure \ref{fig:stoneInPotential} visualises this {\sl ball-in-a-well} approach. For very general non-linear transporters, including the {\sc Michaelis-Menten} form, Eqs.(\ref{eq:MichMent},\ref{eq:MichMente}), used in this paper, the corresponding potential function $V(x)$ displays a single minimum, although its exact shape depends both on the linear and the non-linear terms in Eqs. (\ref{eq:ThreeCompMod}). 

Once this mechanical simile has been realised, an approximation appears quite natural: the potential $V(x)$ is replaced by the parabolic approximation around its minimum, equilibrium, value $x_{eq}$, i.e. approximate  $V(x)\approx V(x_{eq})+\frac{\omega^2}{2}(x-x_{eq})^2$. Similarly, the friction coefficient is replaced by its value at this minimum $\Gamma(x)\approx \Gamma(x_{eq})=\Gamma_{eq}$. With these approximations, {\sc Newton}'s equation (\ref{eq:PotFricGeneral}) becomes the linear equation $\ddot x=-\Gamma_{eq} \dot x-\omega^2(x-x_{eq})$, which describes the damped pendulum of frequency $\omega$. The solution of this equation, as found in many elementary books of mechanics, can be written as the sum of exponential functions of time. One can then undo the change of variables and write the solution in full as done above (eqs. \ref{eq:solution_QaQbQc})

\subsection{Parameters and simplifications for three-compartment system}

The conservation law (\ref{eq:ThreeCompMod4}) allows the elimination of one of the three equations of the set (\ref{eq:ThreeCompMod1}-\ref{eq:ThreeCompMod3}). We have chosen to eliminate either $Q_A$ or $Q_C$, depending on whether the non-linearity depends on $Q_C$ or $Q_A$, respectively. More precisely: if ${\cal J}={\cal J}(Q_C)$, we define the dimensionless normalised concentration $\displaystyle x(t)=\frac{Q_C(t)}{V_C\cal N}$ where ${\cal N}$ is a normalisation constant with units of concentration to be specified later; in the other case, when ${\cal J}={\cal J}(Q_A)$, we define $\displaystyle x(t)=\frac{Q_A(t)}{V_A\cal N}$. In both cases, we also define $\displaystyle y(t)=\frac{Q_B(t)}{V_B\cal N}$. The initial drug concentration $C_0=Q_0/V_I$ (being $V_I=V_A$ or $V_I=V_B$ according to whether the drug is initially loaded on the apical or the basolateral volume, respectively) is also rescaled to $\displaystyle c_0=\frac{C_0}{\cal N}$. We finally define a rescaled dimensionless time variable $\displaystyle s=\frac{Cl_{CB}}{V_B}t$. It turns out that, with these definitions, the resulting two-degree system can be written in the common form:
\begin{subeqnarray}
\label{eq:xy_t}
        \slabel{eq:x_t}
        \dot{x}(s)=\frac{d x(s)}{d s}&=&a_{11} x + a_{12} y + a_{13} + j(x)\\
        \slabel{eq:y_t}
        \dot{y}(s)=\frac{d y(s)}{d s}&=&a_{21} x + a_{22} y + a_{23} .
\end{subeqnarray}
with $\displaystyle j=\frac{V_B}{Cl_{CB}V_A \cal N}\cal J(Q_A)$ in one case, and $\displaystyle j=-\frac{V_B}{Cl_{CB}V_C \cal N}\cal J(Q_C)$ in the other. The dimensionless constants $a_{ij}$  depend on the clearances, volumes, overall concentration, as specified in table \ref{tab:coefficients}.

By differentiating $\dot x(s)$ with respect to $s$ using Eq.(\ref{eq:x_t}) and, in the resulting expression, replacing $\dot y(s)$ by Eq.(\ref{eq:y_t}) and $y(s)$ by its isolation from Eq.(\ref{eq:x_t}), we get the form (\ref{eq:PotFricGeneral}):

\begin{equation}
	\ddot{x}=-\Gamma\left(x\right)\dot{x} + F(x),
\end{equation}
with friction coefficient $\Gamma(x)$ and force $F(x)$ given by:
\begin{equation}
\label{eq:gamma-force}
        \Gamma\left(x\right)=\Gamma_0-j'(x),\,\,\        F(x)=\alpha-\beta x-a_{22} j(x).
\end{equation}
$\Gamma_0$, $\alpha$ and $\beta$ are dimensionless, positively-defined constants, whose relation to the coefficients $a_{ij}$ are also detailed in table \ref{tab:coefficients}.

\subsection{Coefficients in case of {\sc Michaelis-Menten} efflux transporter}
This presentation has so far been very general. In the case of  the {\sc Michaelis-Menten} form for the non-linear efflux transport function ${\cal J}$, formulas (\ref{eq:MichMent}) or (\ref{eq:MichMente}), we identify $K_M$ as a characteristic concentration and simply adopt the normalisation constant ${\cal N}=K_M$. Therefore the current is
\begin{equation}
j(x)=\gamma \frac{x}{1+x}
\end{equation}
where  $\gamma$ is written up in table \ref{tab:coefficients}. The  friction coefficient and force are:
\begin{equation}
\label{eq:force3CompMichMent}
         \Gamma(x)=\Gamma_0-\frac{\gamma}{(1+x)^2},\,\,\,\,\,
         F(x)={\alpha}- {\beta}x- {\gamma}a_{22}\frac{x}{1+x}.
\end{equation}
Note that the initial concentration $C_0$ is contained only in $\alpha$ and that the influence of the efflux transporter is found completely in $\gamma$. $\Gamma_0$, $\alpha$ and $\beta$ are independent of the location of the pump. The values used for numerical calculations are taken from \citep{Bermejo2005} and can be seen in table \ref{tab:parameters}.

The potential from which the force $F(x)=-\frac{dV(x)}{dx}$ derives is:
\begin{equation}
\label{eq:potential}
V(x)=(-\alpha+\gamma a_{22}) x + \frac{\beta}2x^2-\gamma a_{22}\ln (1+x)\,.
\end{equation}

The equilibrium concentration $x_{eq}$ is the minimum of the potential, obtained by setting the force in Eq.(\ref{eq:force3CompMichMent}) equal to zero. In the present case of a {\sc Michaelis-Menten} type non-linearity, the solution is obtained without using any further approximations or simplifications other than stated for the original model:
\begin{equation}
\label{eq:xEquilibrium}
  x_{eq}=\frac{\alpha-\beta-\gamma a_{22}+\sqrt{4\alpha\beta+(\alpha-\beta-\gamma a_{22})^2}}{2\beta}\,.
\end{equation}
Inserted into (\ref{eq:y_t}) the equilibrium value for $y(s)$ is the solution of $\dot y(s)=0$:
\begin{equation}
\label{eq:yEquilibrium}
y_{eq}=-\frac{a_{21}x_{eq}+a_{23}}{a_{22}}
\end{equation}
Through the conservation law one gets $z_{eq}$:
\begin{equation}
\label{eq:zEquilibrium}
z_{eq}=\frac{Q_0}{V_z \cal{N}}-x_{eq}\frac{V_x}{V_z }-y_{eq}\frac{V_B}{V_z}\,,
\end{equation}
where, to unify notation, we have labelled the volumes' meaning $V_x=V_C$ and $V_z=V_A$ for intracellular binding, and  $V_x=V_A$ and $V_z=V_C$  for extracellular binding.

\subsection{The linearised model}
\begin{figure}
 \centering
 \includegraphics[]{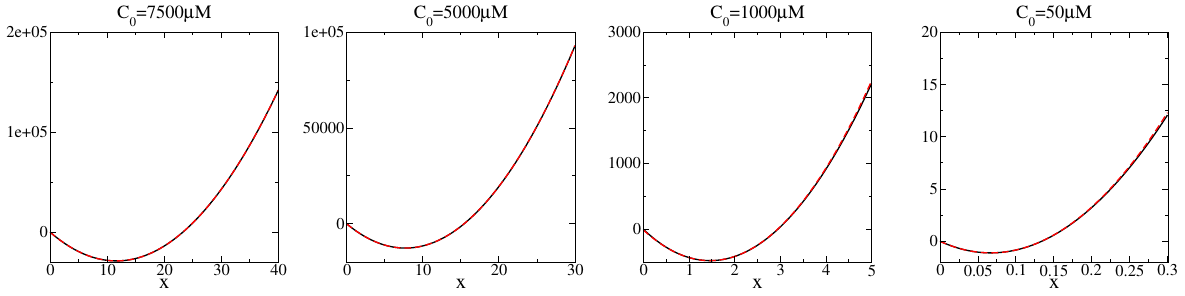}
 \caption{{\it Continuous} line: Original potential Eq.~(\ref{eq:potential}), {\it dashed} line: approximated parabola. For different amounts of loaded drug. Differences are of the order of the line width.}
 \label{fig:potentials}
\end{figure}

The potential $V(x)$ , Eq. (\ref{eq:potential}), has a single minimum and it can be approximated by a parabola around this minimum  $V(x)=V_{eq}+\frac{1}{2}V''_{eq}\left(x-x_{eq}\right)^2$. The second derivative with respect to $x$ is:
\begin{equation}
\label{eq:omega}
V''\left(x_{eq}\right)=\omega^2=\beta+{\frac {{\it a_{22}}\,\gamma}{ \left( 1+{\it x_{eq}} \right) ^{2}}}  \, .
\end{equation}
The approximated and original potentials are drawn in figure \ref{fig:potentials} in the case of apical loading of a system with intracellular binding site. From this figure, it can be seen that the parabolic form constitutes an excellent approximation for a wide range of initial concentrations $C_0$.

Furthermore, we also approximate the friction coefficient $\Gamma(x)$ by its value at equilibrium $\Gamma_{eq}= \Gamma_0-\frac{\gamma}{\left(1+x_{eq}\right)^2}$. The resulting linear differential equation $\ddot x=-\Gamma_{eq} \dot x-\omega^2(x-x_{eq})$ has the solution~\citep{mathsbookBoyceDiPrima}:
\begin{equation}
\label{eq:solDimlessX}
 x(s)=\tilde C_1 e^{-s/\tau_1}+\tilde C_2e^{-s/\tau_2}+x_{eq}
\end{equation}
with time constants:
\begin{equation}
\label{eq:timeconstants}
 \tau_{1}=\frac{2}{\Gamma_{eq}-\sqrt{\Gamma_{eq}^2-4\omega^2}} \,\,\,\,{\rm   and  }\,\,\,\, \tau_{2}=\frac{2}{\Gamma_{eq}+\sqrt{\Gamma_{eq}^2-4\omega^2}}
\end{equation}
The coefficients $\tilde C_{1}$ and $\tilde C_2$ are determined by the initial conditions $x_0$ and $\dot x_0$, the latter being determined through Eq.(\ref{eq:y_t}) by $x_0$ and $y_0$ as  $\dot x_0=a_{11}x_0+a_{12}y_0+a_{13}j(x_0)$. The resulting formulas are summarised in table \ref{tab:coeffsDimless}. The initial values $x_0$ and $y_0$ depend on the particular model used. For example, for intracellular binding we have $x_0=\frac{Q_C(t=0)}{V_C K_M}=0$ and $y_0=\frac{Q_B(t=0)}{V_B K_M}$.  This latter value, in turn, has to be adjusted according to where the drug is loaded: for apical loading $y_0=0$, while for basolateral loading $y_0=\frac{C_0}{K_M}$. Note that $\tau_1$ and $\tau_2$ will be complex if $\Gamma_{eq}<2 \omega$. In this case the system would relax to the steady state by performing damped oscillations. However, for the values of the parameters drawn from experiments, this case does not arise, $\tau_1$ and $\tau_2$ are positive real numbers and the decay to the equilibrium state is governed by real exponentials.

The time evolution of $y$ can be obtained by a direct integration of the linear equation (\ref{eq:y_t}):
\begin{equation}
y(s)=e^{a_{22}s}\left[y_0+\int_0^s ds'\,e^{-a_{22}s'}\left(a_{21}x(s')+a_{23}\right)\right]
\end{equation}
which yields
\begin{equation}
\label{eq:solDimlessY}
 y(s)=\tilde D_1 e^{-s/\tau_1}+\tilde D_2e^{-s/\tau_2}+\tilde D_3 e^{-s/\tau_3}+ y_{eq} \,.
\end{equation}
with a new time constant:
\begin{equation}
\tau_{3}=-\frac{1}{a_{22}}
\end{equation}
and coefficients $\tilde D_{1/2/3}$, whose relation to other constants is detailed in table \ref{tab:coeffsDimless}. The evolution of $z(s)$ is obtained by means of the conservation law. Finally, when we undo the changes of variables we obtain coefficients as given in table \ref{tab:coefficients2} and together with Eq.(\ref{eq:timeconstants}) we derive Eqs. (\ref{eq:solution_QaQbQc}). It is worth recalling that the time scales in real units are 
\begin{equation}
t_1=\frac{V_B}{Cl_{CB}}\tau_1 \,,
t_2=\frac{V_B}{Cl_{CB}}\tau_2 \,,
t_3=\frac{V_B}{Cl_{CB}}\tau_3\,.
\end{equation}

\newpage


\begin{table}
        \caption{Experimental parameters drawn from \citep{Bermejo2005} used for the  calculations.}
        \label{tab:parameters}
        \begin{center}
        \begin{tabular}{|c|c|}
        \hline 
        Parameter &  Measured Value\\
        \hline
        $Cl_{AC} $ & $14.49 \times 10^{-5} cm^3/s$ \\
        $Cl_{CB} $ & $3.528 \times 10^{-3} cm^3/s$ \\
        $V_M$ & $6.17 \times 10^{-12} mol/(cm^2 s)$ \\
        $K_M$ & $0.376 mol/cm^3$ \\
        $S$ & $4.2 cm^2$ \\
        $V_A$ & $2 cm^3$ \\
        $V_B$ & $3 cm^3$ \\
        $V_C$ & $0.0738 cm^3$\\
        \hline
        \end{tabular}
        \end{center}
\end{table}
\begin{table}
       \caption{Coefficients for equations (\ref{eq:xy_t}) using parameter values from \citep{Bermejo2005}.}
       \label{tab:coefficients}
       \begin{center}
       \begin{tabular}{|c|c|c||c|c|}
               \hline
               Parameter &  INTRAcellular     &numerical  & EXTRAcellular &  numerical\\
               \hline
               $a_{11}$ &  $-\left[\frac{V_B}{V_C}+\frac{Cl_{AC}}{Cl_{CB}}\left( \frac{V_B}{V_C}+\frac{V_B}{V_A}\right)\right]$  &-213.8 & $-\frac{Cl_{AC}}{Cl_{CB}}\left(\frac{V_B}{V_C}+\frac{V_B}{V_A}\right)$ & -173.1 \\
               $a_{12}$ &  $\frac{V_B}{V_C}\left(1-\frac{Cl_{AC}}{Cl_{CB}} \frac{V_B}{V_A}\right)$  & -209.8  &  $-\frac{Cl_{AC}}{Cl_{CB}} \frac{V_B^2}{V_AV_C}$& -250.4 \\
               $a_{13}$ & $\frac{Cl_{AC}}{Cl_{CB}} \frac{V_BV_I}{V_AV_C}c_0$  & 2.220$\times 10^8 V_I C_0$ &  $\frac{Cl_{AC}}{Cl_{CB}} \frac{V_BV_I}{V_AV_C}c_0$ & 2.220$\times 10^8 V_I C_0$  \\
               $a_{21}$ &  $1$  & 1 &  $-\frac{V_A}{V_C}$& -27.10\\
               $a_{22}$ &  $-1$  & -1&  $-\left(\frac{V_B}{V_C}+1\right)$& -41.65\\
               $a_{23}$&  $0$   & 0 &  $c_0 \frac{V_I}{V_C}$& 3.604$\times 10^8 V_I C_0$ \\
 	       	$\gamma$ & $-\frac{SV_M}{Cl_{CB}K_M}\frac{V_B}{V_C}$ & -79.41 & $\frac{SV_M}{Cl_{CB}K_M}\frac{V_B}{V_A}$& 2.930 \\
\hline
	     \end{tabular}
\newline
\newline
\newline
\newline
       \begin{tabular}{|c|rcl|c|}
       \hline
	     	       $\Gamma_0$ & $-a_{11}-a_{22}$& 	= & $\frac{V_B}{V_C}+\frac{Cl_{AC}}{Cl_{CB}}\left( \frac{V_B}{V_C}+\frac{V_B}{V_A}\right)+1$& 	214.8 \\
	       $\alpha$ & $a_{12}a_{23}-a_{13}a_{22}$ & = & $\frac{Cl_{AC}}{Cl_{CB}}  \frac{V_BV_I}{V_AV_C}c_0$ & 2.220$\times 10^8 V_I C_0$ \\
	       $\beta$ & $a_{11}a_{22}-a_{12}a_{21}$ & = & $\frac{Cl_{AC}}{Cl_{CB}}\frac{V_B}{V_AV_C}\left(V_A+V_B+V_C\right)$& 423.6 \\
               \hline
       \end{tabular}
       \end{center}
\end{table}
\begin{table}
\caption{Coefficients of solutions (\ref{eq:solution_QaQbQc}). For equilibrium solution $x_{eq},y_{eq}$ and coefficients $\tilde C_i$ and $\tilde D_i$ see text in appendix and table \ref{tab:coeffsDimless}. The numerical values are calculated for $C_0=7500\mu M$ loaded apically.}
	\label{tab:coefficients2}
\begin{center}
\begin{tabular}{|c|c|c||c|c|}
\hline
& INTRA &numerical ($\mu mol$)& EXTRA &numerical ($\mu mol$)\\
\hline
$A_1$ & $K_M  \left( V_C \tilde C_1+V_B \tilde D_1 \right)$&$-8.90$  & $-K_M  V_A \tilde C_1$&$-5.74$\\
$A_2$ & $K_M  \left( V_C \tilde C_2+V_B \tilde D_2 \right)$&$-0.341$ & $-K_M  V_A \tilde C_2$&$-11.8$ \\
$A_3$ & $K_M V_B\tilde D_3$                                &$0.348$&  $0$  &  $0$\\
$Q_A^{eq}$ &  $Q_0-K_M \left(V_Cx_{eq}+ V_By_{eq}\right)$  &$6.1045$ & $K_M V_Ax_{eq}$&$6.1057$\\
$B_1$ & $-K_M  V_B \tilde D_1$                             &$9.11\,$           & $-K_M  V_B \tilde D_1$ &$ 5.88$ \\
$B_2$ & $-K_M  V_B \tilde D_2$                             &$-0.0769$ & $-K_M  V_B \tilde D_2$ & $2.82$ \\
$B_3$ &$-K_M  V_B \tilde D_3$                              &$-0.348$ & $-K_M  V_B \tilde D_3$ & $-0.0153$ \\
$Q_B^{eq}$ & $K_M  V_B\,y_{eq}$                            &$8.6819\,$ & $K_M  V_B\,y_{eq}$ & $8.6808$ \\
$C_1$ &$-K_M  V_C \tilde C_1 $                             &$ -0.204$ & $K_M \left( V_A \tilde C_1+V_B\tilde D_1\right) $&$ -0.143$ \\
$C_2$ &$-K_M  V_C \tilde C_2 $                             &$ 0.418$ & $K_M \left( V_A \tilde C_2+V_B\tilde D_2\right)  $&$ -14.7$ \\
$C_3$ &  $0$&$0$ 			  & $K_M  V_B \tilde D_3 $&$ 0.0153$\\
$Q_C^{eq}$ & $K_M  V_C x_{eq} $&$ 0.21358$&$Q_0-K_M \left(V_A x_{eq}+V_B y_{eq}\right) $&$ 0.21355$\\
\hline
\end{tabular}
\end{center}
\end{table}
\newpage
\begin{table}[ht]
	\caption{Coefficients of dimensionless solution eqs. (\ref{eq:solDimlessX}) and (\ref{eq:solDimlessY}).}
	\label{tab:coeffsDimless}
\begin{center}
\begin{tabular}{|c|c|}
\hline
$\tilde C_1$&$ \left(x_0-x_{eq}+\tau_2\left(a_{11}x_0+a_{12}y_0+a_{13}+\gamma\frac{x_0}{1+x_0}\right)\right)\frac{\tau_1}{\tau_1-\tau_2}$\\
$\tilde C_2$&$\displaystyle x_0-x_{eq}-\tilde C_1$\\
$\tilde D_1$&$\displaystyle -\frac{a_{21}\tilde C_1\tau_1}{a_{22}\tau_1+1}$\\
$\tilde D_2$&$\displaystyle -\frac{a_{21}\tilde C_2\tau_2}{a_{22}\tau_2+1}$\\
$\tilde D_3$&$\displaystyle y_0-y_{eq} -\tilde D_1-\tilde D_2$\\
\hline
\end{tabular}
\end{center}
\end{table}

\newpage
\bibliographystyle{elsart-harv}
\bibliography{bibliography}
\end{document}